\documentclass[11pt,letterpaper]{article}

\usepackage[margin=1in]{geometry}
\usepackage[utf8]{inputenc}
\usepackage{microtype}
\usepackage{amsmath,amssymb}
\usepackage{booktabs}
\usepackage{graphicx}
\usepackage{adjustbox}
\usepackage{caption}
\usepackage{subcaption}
\usepackage{enumitem}
\usepackage{xcolor}
\usepackage{tikz}
\usetikzlibrary{arrows.meta,positioning,shapes.misc}
\usepackage{setspace}
\usepackage{multicol}
\usepackage{float}
\PassOptionsToPackage{hyphens}{url}
\usepackage[authoryear,round]{natbib}
\usepackage[hidelinks]{hyperref}
\usepackage{url}
\setcitestyle{aysep={,}}

\graphicspath{{figs/}}
\setlist{nosep}
\setlist[itemize,1]{label=\ensuremath{\bullet}}
\definecolor{glblue}{HTML}{2A78D6}
\definecolor{glorange}{HTML}{EB6834}
\definecolor{glgray}{HTML}{8A8985}
\definecolor{glink}{HTML}{0B0B0B}

\title{\textbf{Match Rate Is Not Match Quality}\\[4pt]
\large An Empirical Assessment of Third-Party Identity Resolution\\
Using Random Device Engagement Polling and a National Voter File}
\author{Tobias B. Konitzer\thanks{GrowthLoop. Email: \texttt{tobias.konitzer@gmail.com}} \and David M. Rothschild\thanks{Email: \texttt{david@researchdmr.com}}}

\date{\today}

\begin{document}
\maketitle

\begin{abstract}
\noindent Third-party identity-resolution vendors commonly emphasize \emph{match rate}, the share of submitted records they can link to identifiers for advertising activation, as their core metric of quality. Yet a high match rate does not necessarily mean that those matched identifiers belong to the submitted records (i.e., the intended targets). In this paper we explore \emph{match quality} as an alternative metric to match rate, utilizing a unique data set to evaluate match quality. We are able to compare two demographics, age and gender, from a submitted record (voter files) to the same demographics in self-reported surveys matched to mobile advertising identifiers linked together by two leading identity-resolution vendors. The data and matching was from 2019-20. There is substantial ambiguity in match quality of the returned records. First, Vendor A's activation files contain an average of 3.4 device identifiers per match, while Vendor B returns an average of 7.9 voter records per match. Some people own multiple devices that each have unique mobile identifiers and some people may share devices within their household, but with this ambiguity a headline match rate does not reveal whether the returned links identify the intended person, another household member, or someone else. Fortunately, Vendor B explicitly labels some matches as individual-level rather than household-level matches, yet among those, age bracket and gender agree in 23.6\% of pairs. Vendor A has a similar match quality, with agreement of 21.8\% across audited device--record links. Utilizing a mixture model that assumes that correct links always agree on these two demographics and incorrect links agree 10\% of the time, these rates imply conditional identity-level precisions of approximately 15\% and 13\%, respectively. This model makes assumptions about demographic measurement error and chance agreement among incorrect links that we prove robust in the paper. This paper utilizes identity-resolution products that are several years old, but the question we address remains fundamental: match rate alone cannot establish whether advertisers are reaching the people they intended. Evaluating identity resolution requires independent evidence of match quality alongside match rate.
\end{abstract}

\section{Introduction}
Since the dawn of online advertising, identity resolution has been at the forefront of discussions about reach, quality of the ad inventory and, ultimately, at the heart of assessments whether ad targeting has been successful or not. This is particularly true for audience-based advertising or addressable advertising that starts with a known, but more often than not anonymous, user-level identifier. According to Yahoo Advertising Platform Insights, the match rate is a ``well-established metric that serves as an indicator to gauge how well an audience can scale within a platform'' \citep{yahoo_matchrate}. While advertisers at times seek broad targeting with the goal of purchase maximization, reaching a specific set of users remains a large use case.

Thus, identity resolution is also a key offering of vendors in the advertising space. In this context, companies (i.e. the advertiser) send CRM (Customer Relationship Management, or Customer 360) records to third-party identity resolution vendors, who oftentimes directly onboard these identities to third-party ad platforms, such as Meta. The core value proposition of many if not most identity resolution vendors is that activating a set of known or unknown IDs within walled gardens such as Meta will receive a significant boost in match-rate if the vendor's identity spine is used \citep[see, e.g.,][]{liveramp_capi}. These vendors will frequently focus on match rate as the key metric of success.

We argue that match rate alone is a misleading success metric and turn our attention to the \emph{match quality}. If the campaign ultimately targets a set of users that is in fact \emph{not} part of the subset of users exported from the generator (e.g., an e-commerce website trying to target its customers via online advertising), match rate may be high, but the intention behind the targeting is violated.

Quality of the match instead asks: to what extent are the identities that the identity vendor exports to the target destination after processing through its identity spine \emph{the same identities} that the generator initially shared with the vendor? Given industry standards, vendors may have an incentive to maximize match rate \emph{even if the quality of the match} is affected adversely.

Usually, quality of the match cannot be assessed at an individual level, because target destinations do not share back individual-level information on identities they receive, and most vendors strip data of PII fields before exporting to target destinations.\footnote{Here is how a leading identity-resolution vendor discusses hashed IDs: \url{https://liveramp.com/blog/powering-better-conversions-api-capi-results-with-rampid}} For example, it is theoretically possible that ad platforms cannot reach an intended audience, but serve the ads to a similar set of targets that overlaps with the initial audience on all demographics the ad network reports on, resulting in a match accuracy of $0$, entirely invisible and inauditable to the generator. This may be ok if the goal is not tied to reaching \textit{specific} targets, but to maximize some prespecified return (such as revenue), but in many other applications, including retargeting etc., the intent of generator is to target the same individuals it submitted to the identity resolution vendor.

\subsection*{Match accuracy and Match rate -- an illustration}
\begin{figure}[H]
\centering
\includegraphics[width=\textwidth]{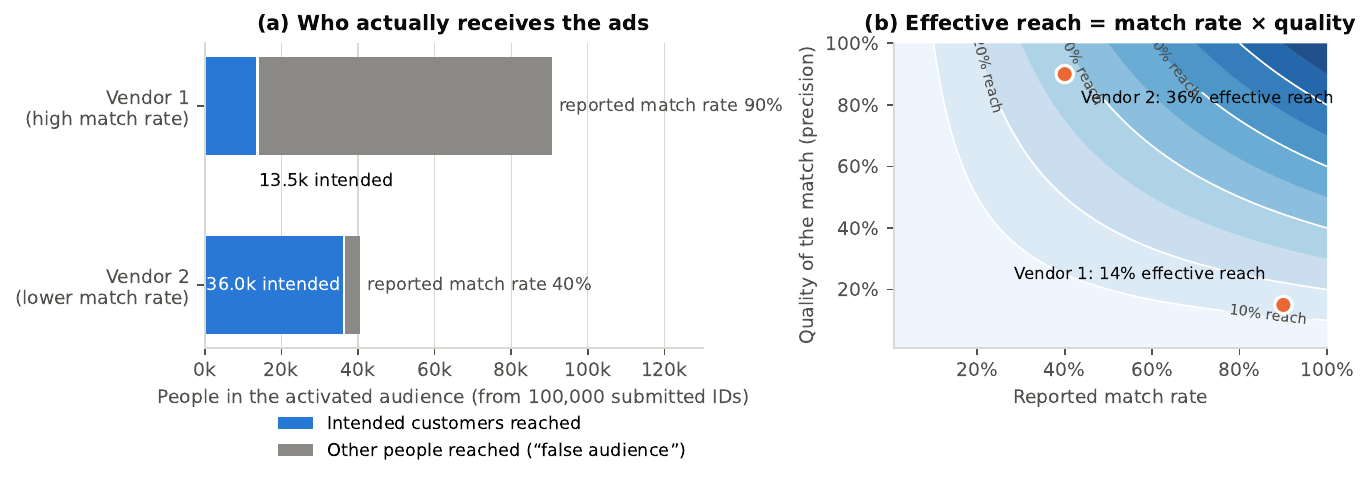}
\caption{A stylized example of high match rate with low match quality. \textbf{(a)} Composition of the activated audience for two hypothetical vendors given 100{,}000 submitted IDs; the vendor with the higher reported match rate reaches fewer of the intended customers and many more bystanders. \textbf{(b)} Effective reach (share of the submitted list activated \emph{and} correctly identified) as the product of match rate and quality; iso-reach curves show that very different match rates can deliver identical effective reach.}
\label{fig:theory}
\end{figure}

Consider an e-commerce retailer that wants to export $N = 100{,}000$ customer IDs from their CRM for activation on a third-party ad platform. Two identity vendors bid for the service. Vendor 1 reports a match rate of $m_H = 90\%$; Vendor 2 reports only $m_L = 40\%$. Vendor 1 wins the industry-standard assessment based on match rate only by a wide margin. Let us now consider the \emph{quality} of the match, $q$, defined as the probability that an activated identifier belongs to the very person whose ID was submitted, i.e. individual-level congruence (i.e. the quality of the match). Suppose Vendor 1 achieves its high match rate by allowing matches with very low confidence that include spurious household-level, probabilistic, and single-touchpoint links, etc., so that $q_H = 15\%$, whereas Vendor 2 only returns individual-level, deterministic, multi-touchpoint links and achieves $q_L = 90\%$. For this stylized illustration, assume that each matched submitted record produces one distinct activated person, with no duplication, and that incorrect links point to people outside the submitted customer list. The implied activated audiences are then:
\begin{align*}
\text{Vendor 1:}\quad & 90{,}000 \text{ activated} = \underbrace{13{,}500}_{\text{intended customers}} + \underbrace{76{,}500}_{\text{other people}},\\
\text{Vendor 2:}\quad & 40{,}000 \text{ activated} = \underbrace{36{,}000}_{\text{intended customers}} + \underbrace{\phantom{7}4{,}000}_{\text{other people}}.
\end{align*}
Vendor 1's activated audience is more than twice as large, yet under these assumptions it contains fewer than half as many of the retailer's actual customers, and 85\% of its activated audience falls outside the submitted customer list (Figure~\ref{fig:theory}a). The relevant quantity in this illustration is not match rate $m$ alone but $m\times q$, the share of submitted records that are both activated and correctly linked (Figure~\ref{fig:theory}b).

This paper uses a rare opportunity to assess match quality for two leading identity-resolution vendors using data collected in 2019--2020. Comparing survey respondents' self-reported age and gender with the attributes of their linked voter-file records, we find demographic agreement only modestly above a random-assignment benchmark for one vendor and higher, but still low, agreement for the other. For Vendor B, our primary assessment focuses on pairs the vendor labels as individual-level matches. Under a mixture model that accounts for chance demographic agreement, these pairs and Vendor A's audited links imply conditional identity-level precisions of approximately 15\% and 13\%, respectively. Although these findings do not establish the performance of today's products, they demonstrate why match rate alone is insufficient to evaluate identity resolution and why independent validation of match quality is necessary. We first describe the matching exercises and the structure of the returned links, then assess demographic agreement and its implications for identity-level precision.

\section{A rare opportunity to assess match quality}
Match quality is usually difficult for advertisers to assess independently because they generally cannot link users reached on an ad platform back to the individual records they submitted. Ad networks do not share individual identifiers, or other identifiable information, of targeted users other than coarse demographic breakdowns.

In 2019--2020 now defunct survey research company \textit{PredictWise} used two leading identity-resolution vendors to link voter file data to online identifiers (Mobile Ad IDs, or MAIDs, common meta-data on surveys at the time): Vendor A (LiveRamp) matched voter file data to MAIDs through its PII-based identity graph; Vendor B (Kochava), a mobile measurement company, matched a set of MAIDs that PredictWise had recently seen in the field to voter file records through its own PII-based device-ID product.\footnote{The voter file is data on voter turnout and demographics, curated by the states and enhanced and homogenized by commercial voter-file vendors; on its use as a validation source see \citet{ansolabehere2012validation} and \citet{hersh2015hacking}. MAIDs are semi-permanent identifiers identifying users of mobile devices across applications, ad networks, exchanges and analytics platforms while preserving PII-based privacy of these users. Google/Android and Apple/iOS assign MAIDs to holders of mobile devices as AAIDs and IDFAs respectively. Apple/iOS users can opt out of ad tracking entirely, meaning that no permanent IDFA will be assigned to these users. Google/Android users can merely reset AAIDs, but not opt out of MAID assignments entirely. Further note that both authors were associated with PredictWise.} PredictWise then targeted these MAIDs with additional in-app survey requests on the phones tapping into the in-app ad space.\footnote {see \url{https://www.greenbook.org/insights/research-methodologies/the-methodology-that-is-overcoming-some-of-market-researchs-biggest-obstacles} for the surveying methodology} Here, the voter file is analogous to CRM/customer 360, and the in-app surveying enhanced with the match is analogous to identity-enhanced ad targeting. Both matching directions allow us to evaluate demographic agreement within returned record--device pairs, although their match rates and link structures are not directly comparable. In contrast to more traditional ad networks, tapping into the in-app ad space with survey requests allows us to receive back the MAID of the sampled individual as telemetry, along with the survey results.

Thus, this data allows us to assess match quality directly for two leading vendors in the space (at the time), and in the case of Vendor A for the vendor's flagship data product at the time: The trade publication AdExchanger noted that ``as a data onboarder, [their] bread and butter has been to match a marketer's offline customer data with pseudonymous digital identifiers for the purposes of analytics or activation'' \citep{adexchanger_liveramp}.\footnote{We should note that while MAID-to-PII matching was the identity-resolution vendor's historical stronghold, 2021, shortly after our data were collected, was a period of intense structural pivot due to major industry privacy changes: Apple's iOS 14.5 App Tracking Transparency (ATT) framework was released in April 2021 and required users to opt in to tracking, causing IDFA availability to plummet overnight. Announcement is here: \url{https://developer.apple.com/news/?id=ecvrtzt2}.} In our set-up we can simply cross-tabulate self-reported demographics obtained by the follow-up surveys overlapping with the matched voter file data, particularly age and gender.\footnote{Demographic disagreement can reflect incorrect identity links, but also inaccuracies in the voter file, survey response error, or changes to demographics between measurement dates. We therefore do not consider demographic disagreement as a direct verification of identity mismatch, but rather a data point towards understanding match quality; the subsequent identity-precision estimates depend on assumptions about these measurement errors.}

\section{An empirical investigation of the match rate claim}\label{sec:structure}

Like many other vendors in the space, Vendor A claims to boost match rates by greater than 50\% over baselines such as hashed email to hashed email.\footnote{\url{https://liveramp.com/blog/powering-better-conversions-api-capi-results-with-rampid}} Vendor B likewise claims to boost match rates by individual-level, deterministic matching of offline to online identities.\footnote{See \url{https://support.kochava.com/articles/reference-information/2649-matched-by-definitions/}} The vendors are no exception to the mechanism described above, in that they ingest first-party data, for example from brands that want to advertise on Meta etc., strip the data of Personal Identifiable Information (PII), enhance the data with their proprietary identity match, and export that data to Meta directly. 

Note that possible one-(voter file record)-to-many (MAID) and many-(voter file records)-to-one (MAID) matches complicate our ensuing analysis. One-(voter file record)-to-many (MAID) matches can occur for several reasons. (a)~Users are active on iOS-based devices (such as an iPad) and Android-based devices (such as a Samsung Galaxy) at the same time; (b)~users maintain different accounts within Google services or within iCloud/Apple services;\footnote{For example, one identity link is tied to 300 different IDFAs and 300 different AAIDs.} (c)~matches come with different matching quality: for example, the identity resolution vendor holds matches from \emph{maintained} records, i.e.\ from individuals that they fully recognize via multiple PII touch-points to multiple digital footprints; \emph{derived} records, i.e.\ records identified by a simple PII touchpoint; and matches at the household level. 

Many-(voter file records)-to-one (MAID) matches are more problematic. These many-to-one matches can occur for several reasons: (a)~people share devices, especially tablets and other household devices (Table~\ref{tab:sharing}); (b)~vendor identifies records on the voter file as duplicates; (c)~matches are done at the household level; (d)~or matches are erroneous. Many-to-one matches bring with them the danger of targeting the wrong individual, and related boomerang effects. Note that the ``household problem''---resolving online or other identifiers to individuals within the \emph{same} household---is a known and widely studied problem in record linkage \citep[see, e.g.,][]{fu2014household}.   

\begin{table}[H]
\centering
\small
\begin{tabular}{lcccc}
\toprule
\textbf{MAID} & \textbf{Age} & \textbf{Gender} & \textbf{Census block} & \textbf{State} \\
\midrule
001b19e1-a11d-41d2-\#\#\# & 72 & Female & 450570112012020 & SC \\
001b19e1-a11d-41d2-\#\#\# & 68 & Female & 450570112012020 & SC \\
\bottomrule
\end{tabular}
\caption{Possible device sharing among two independent records on the voter file.}
\label{tab:sharing}
\end{table}

\begin{figure}[H]
\centering
\adjustbox{trim={0 0 {.475\width} 0},clip,width=0.58\textwidth}{\includegraphics{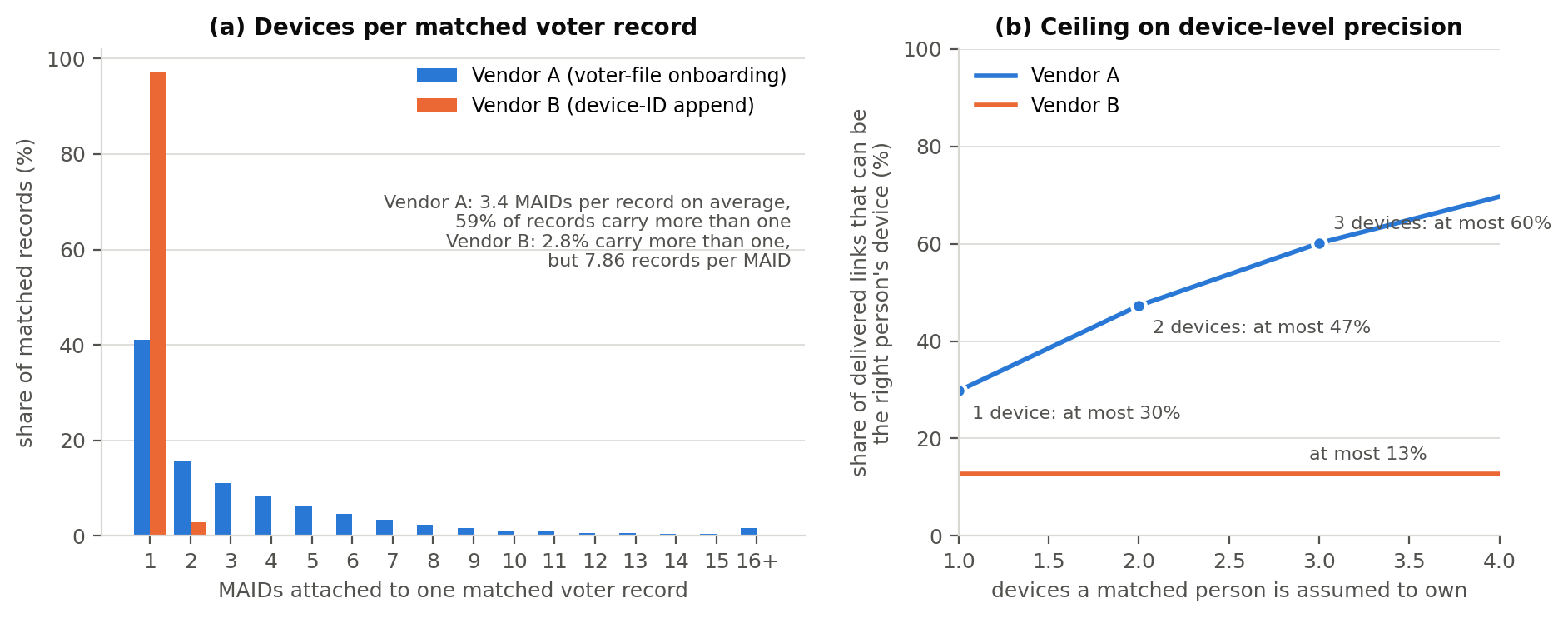}}
\caption{Devices per matched voter record. Matched voter records by the number of MAIDs attached to them. Vendor A: the five state activation files described above. Vendor B: the 1,176 matched MAIDs; almost every record carries one MAID, but each MAID carries 7.86 records on average.}
\label{fig:maidstructure}
\end{figure}

Both vendors return one-to-many and many-to-one links (Figure~\ref{fig:maidstructure}). Vendor A returned MAIDs for 65.34\% of submitted voter-file records (match rate is where most evaluations end, but here is where we begin). We examine five state activation files (Arizona, Florida, Michigan, North Carolina, and Pennsylvania), containing 2.29 million voter records and 7.68 million MAIDs after many-to-one links were resolved to a single voter record within each state by the vendor. Each voter record carries 3.4 MAIDs on average; 59\% carry more than one, and 24\% of MAIDs are attached to records carrying ten or more.

For demographic validation, we intersect these files with 19{,}154 unique MAIDs in our survey archive for the five states. The overlap contains 117 devices, two linked to different voters across states, yielding 119 device--record links. We retain each device's most recent survey response. These 119 links are the sample we use for the audit of Vendor A. To be included in this sample a device needs to appear in both the activated voter file and the survey archive, thus this is not a representative sample of all links returned by Vendor A. Since both the matching direction and the sample frame differ between the studies of the two vendors, we cannot make a head-to-head comparison. Instead, as both studies reveal substantial similar demographic disagreement, despite the differences between the studies, we view this a confirming set of studies.

For Vendor B, we conduct this analysis using a set of Mobile Ad IDs that had been consistent responders to PredictWise surveys conducted prior to this analysis.\footnote{Arguably, this makes findings of low match accuracy even more conservative, because we are relying on a subset of well-known, easy-to-reach MAIDs for which match accuracy should be highest.} Out of the set of frequently surveyed MAIDs, a random sample of 6,765 was submitted to Vendor B for matching; 1,176 of these were appended to the voter file data by the vendor, for a total match rate of 17.38\%. Again note that this is where most experiences end: the generator settles for knowing 17.38\% of their customers could be matched.

Once more, the returned data includes both link types. One-(voter file)-to-many-(MAID) links are rare: 2.82\% of matched voter-file records are tied to more than one MAID, 2.01 on average. Many-(voter file)-to-one-(MAID) links dominate: 97.03\% of matched MAIDs are tied to more than one voter-file record, and the average matched MAID is tied to 7.86 records, which is the signature of matching that resolves a device to a household rather than to an individual (Figure~\ref{fig:maidstructure}). With 1,176 matched MAIDs at 7.86 records each on average, this yielded roughly 9,200 MAID–voter-record pairs, and accuracy below is computed on these pairs. 

These link structures underscore the distinction between returning an association and identifying a particular individual. Vendor B labels returned pairs as individual-level or household-level matches and also provides confidence codes. Our primary assessment of Vendor B's individual-level match quality therefore focuses on the pairs carrying its individual-level designation; household-level and pooled results describe the broader returned file. For Vendor A, we also examine agreement among records carrying a single MAID within the processed activation files, without treating these as necessarily unambiguous original vendor matches.

Based on the voter-file-ID$\leftrightarrow$MAID pairs returned by the identity-resolution vendor, we can now assess match quality among matches by comparing voter-file-based hard demographics, specifically gender and age, to demographic data collected within the surveys as self-reports, bar clerical error on the side of the voter file and response error on the side of the surveys (as discussed above).

\section{Match Accuracy}

We assess demographic agreement by comparing survey respondents' self-reported age bracket and gender with the corresponding attributes of their linked voter-file records. Age is grouped into five categories (18--24, 25--34, 35--44, 45--54, and 55+), and gender into two categories. Agreement on both attributes is consistent with a correct identity link but does not establish one, because different people can share both attributes. Conversely, disagreement can reflect an incorrect link or demographic measurement error.

\subsection{Vendor A match accuracy}
Among the 119 device--record links in Vendor A's audit sample, gender agrees in 61.3\% of links, age bracket in 32.8\%, and there is joint demographic agreement in just 21.8\% of cases (95\% interval 15--30\%). These results describe the overlap between the five-state activation files and our survey archive, rather than a random sample of all Vendor A links.

There are substantial differences in the match agreement depending on whether or not the states match. The states differ in 57\% of the audited links; joint demographic agreement is just 7.4\% among the links where states do not match. But  there is 41\% joint demographic agreement when states do match. We do not use state as part of the joint demographic comparison because different states do not, by itself, determine the match is incorrect. The same people can easily be in different states over time, especially as most voter do not immediately update their records. In contrast age bracket and gender do not face that same issue. Therefore we report geographic split as descriptive, rather than as an independently validated identity error.

\begin{figure}[H]
\centering
\includegraphics[width=\textwidth]{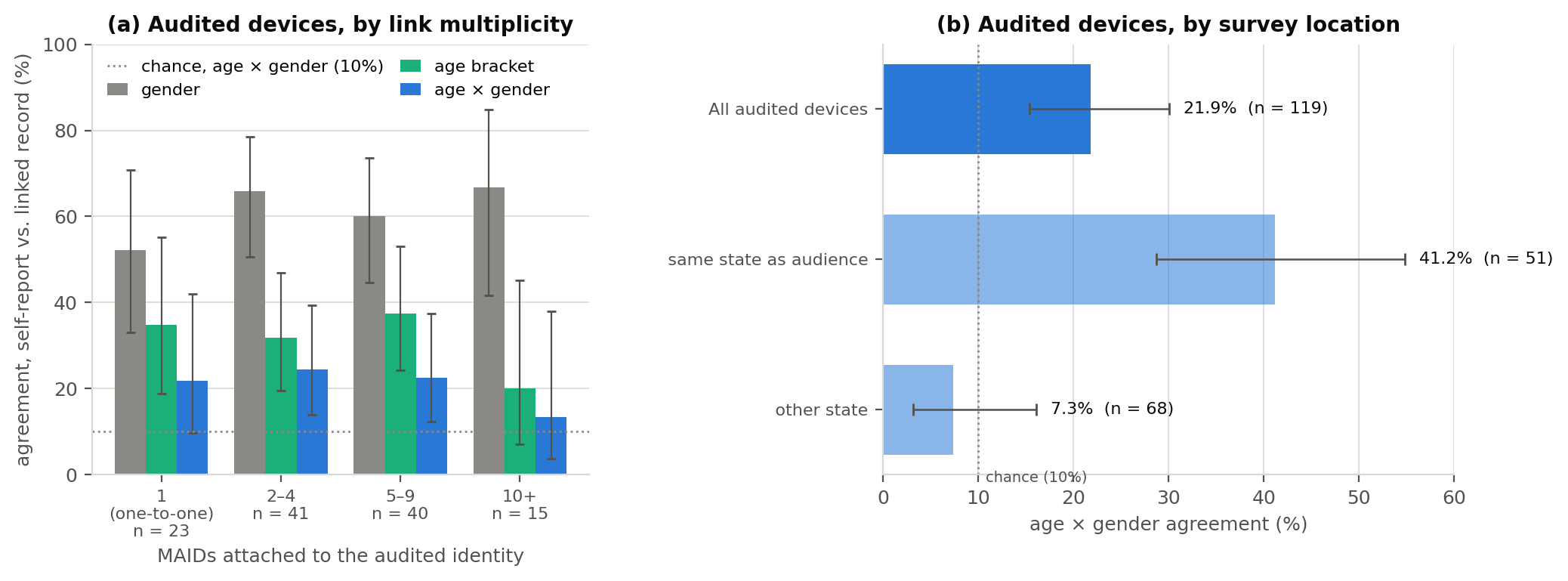}
\caption{Vendor A demographic agreement in the audited overlap ($n = 119$ device--record links). \textbf{(a)} Gender, age-bracket, and joint agreement by the number of MAIDs attached to the linked voter record within the activation file; intervals are 95\% Wilson intervals. \textbf{(b)} Joint agreement overall and by whether the recorded survey state matches the audience state.}
\label{fig:vendorA}
\end{figure}

Figure~\ref{fig:vendorA} also compares agreement across records carrying different numbers of MAIDs. Among the 23 audited links whose voter record carries a single MAID within the activation file, joint agreement is 21.7\% (95\% interval 10--42\%). Corresponding rates are 24.4\% for records carrying 2--4 MAIDs, 22.5\% for those carrying 5--9, and 13.3\% for those carrying ten or more. Agreement is therefore low even among single-MAID records, although the small subgroup samples do not support precise comparisons across multiplicity categories. A single-MAID record within these processed activation files should not be interpreted as an originally unambiguous one-to-one vendor match.

\subsection{Vendor B match accuracy}
\begin{figure}[H]
\centering
\includegraphics[width=\textwidth]{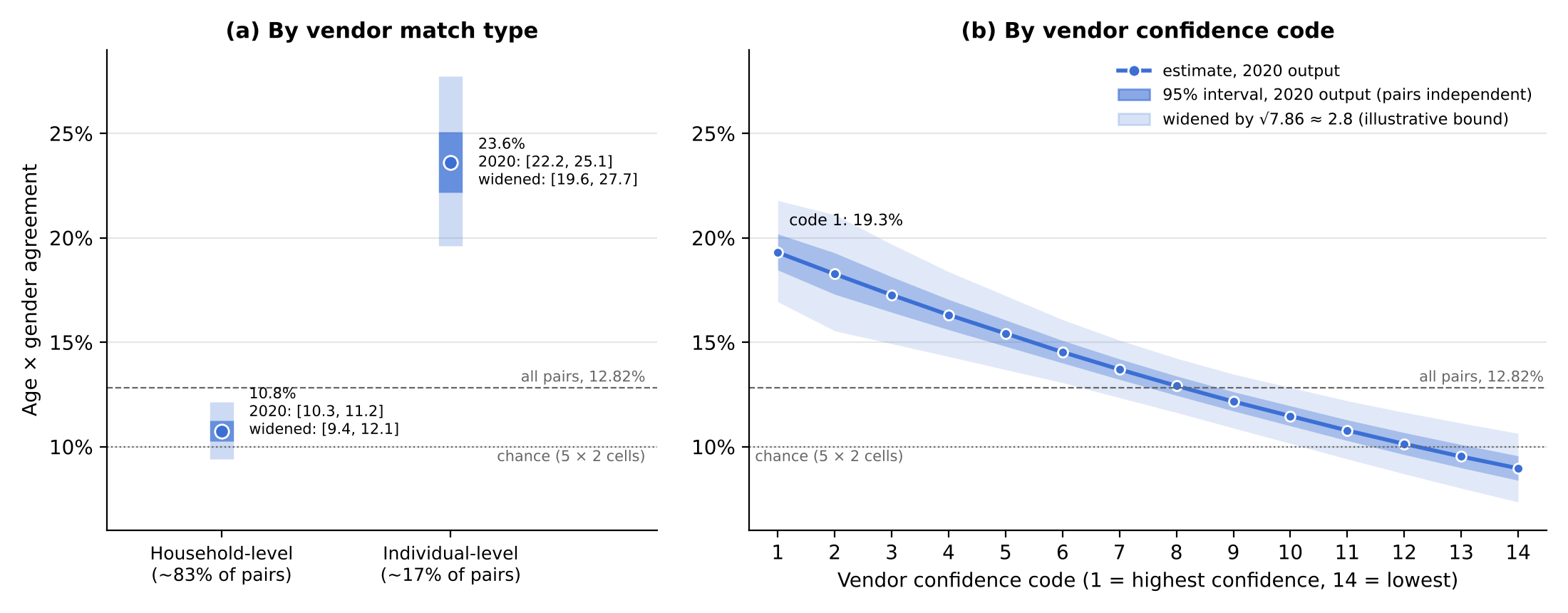}
\caption{Vendor B demographic agreement between voter-file attributes and survey self-reports. \textbf{(a)} Joint age-bracket and gender agreement by vendor-reported match type. \textbf{(b)} Regression-predicted agreement by vendor-reported confidence code; code 1 denotes the highest vendor confidence (agreement rate regressed on confidence code). In both panels, the darker interval is the 95\% interval, which treats record--device pairs as independent. The lighter interval widens it by a $\sqrt{7.86}\approx 2.8$ correction factor, based on the average of 7.86 records per MAID across all pairs and applied to every group. Because the number of records per device varies, and is not known separately by match type or confidence code, this is an illustrative bound, not a cluster-robust estimate.}
\label{fig:regression}
\end{figure}

Vendor B labels each returned record--device pair as an individual-level or household-level match. We focus on the individual-level as our primary assessment because this designation means that the linked record for the survey respondent should be the exact person, not a household member. Within this subset, self-reported survey results and voter-file agree jointly on age bracket and gender in only 23.6\% of pairs (Figure~\ref{fig:regression}a).

For comparison, joint agreement is 10.8\% among household-level pairs and 12.82\% across all returned pairs. These pooled results describe the broader file, not solely the vendor's individual-level links. All agreement rates are pair-level measures: they do not indicate how often at least one of the multiple records returned for a device identifies the respondent correctly.

Figure~\ref{fig:regression}b presents regression-predicted agreement by vendor-reported confidence code across the returned pairs, with code 1 denoting the highest confidence. Because multiple pairs share the same device and survey response, uncertainty estimates should account for clustering within devices. The intervals in Figure~\ref{fig:regression} are therefore  widened by $\sqrt{7.86}\approx 2.8$, the average number of voter records per device, as an illustrative bound rather than an estimate.

For context, uniform random assignment across the ten age--gender categories (i.e., two genders cross five age brackets) would produce 10\% agreement. Random pairing based on the actual demographic distributions need not produce this rate; the empirical benchmark and sensitivity to alternative wrong-person agreement rates are examined in Section~\ref{sec:simulation}.

Overall, the identity-matched data shows meaningful demographic disagreement for both vendors. Vendor B's data which is designated at individual-level pairs, has higher agreement than the rest of Vendor B's data, but even this group has joint demographic agreement in fewer than one-quarter of pairs. The expectation agreement on gender and age bracket is not perfect, with both clerical and response error, but the revealed quality does not approach 100\%. For Vendor A we can isolate the pairs with just one device per person, and even there the agreement remains low. And while it is reassuring to see a higher match rate for Vendor A among those pairs who match in state, this further highlights the heterogeneous quality within the audited sample. We next examine what these agreement rates imply for identity-level precision under explicit assumptions about chance agreement and measurement error.

\section{What the agreement rates imply: a conditional illustration}\label{sec:simulation}

The preceding analysis documents low demographic agreement in the audited links returned by both vendors. To illustrate what these agreement rates could imply for identity-level precision, we use a simple mixture model. The resulting figures are conditional on assumptions about chance agreement and measurement error; they are not estimates of true identity accuracy or of campaign delivery.

\subsection{From demographic agreement to identity-level precision}

The agreement rates quoted above---23.6\% for Vendor B's individual-level pairs and 21.8\% for Vendor A's audited links---are agreement rates on a $5\times2$ demographic grid, not the share of matches that identify the correct person. The two are linked by a simple mixture: a correct match agrees on age and gender essentially always (bar the small clerical and response errors noted above), whereas a wrong-person match agrees with some baseline probability $b$. Essentially, this is the false positive rate: A match can record as ``correct" on our $5\times2$ grid, even though the match point to different identities, for example if a 20-year-old male and a 21-year-old male are matched, our analysis of match quality would incorrectly record a ``correct" match.

If $\pi$ denotes the share of delivered matches that are the intended individual, the observed agreement rate $a$ is:
\begin{equation}
a = \pi + (1-\pi)\,b \qquad\Longrightarrow\qquad \pi = \frac{a-b}{1-b}.
\label{eq:mixture}
\end{equation}
If wrong-person links agreed at the uniform chance rate ($b=1/10$), Vendor B's individual-level pairs ($a=23.6\%$) would imply $\pi\approx15.1\%$. For context, applying the same calculation to all Vendor B pairs, including household-level associations ($a=12.82\%$), yields $\pi\approx3.1\%$.\footnote{We do not report an interval for Vendor B’s implied precision. Any such interval would reflect only sampling error in the observed agreement rate, not the chance rate $b$ nor measurement error $e$.} Applying the same correction to the vendor's own confidence tiers, a confidence-code-1 match ($a \approx 19\%$) would correspond to the right person about 10\% of the time, and even a generous 25\% agreement rate for Vendor B's ``individual-level'' matches (above the 23.6\% observed) would correspond to at most one correct identity in six. For Vendor A, the audited agreement of 21.8\% implies $\pi \approx 13\%$).

To assess sensitivity to demographic measurement error (i.e., the possibility that the same person for some reason had different answers for demographics between data), let $e$ denote
the probability that a correctly linked survey respondent and voter-file
record disagree on age bracket, gender, or both. The mixture then becomes
\begin{equation}
a = \pi(1-e) + (1-\pi)b
\qquad\Longrightarrow\qquad
\pi = \frac{a-b}{(1-e)-b}.
\label{eq:mixture_error}
\end{equation}
Equation~\eqref{eq:mixture} is the special case $e=0$.
Table~\ref{tab:measurement_error} varies $e$ while holding
wrong-person agreement at $b=10\%$. These are sensitivity scenarios,
not estimated measurement-error rates. Allowing 10\% demographic
disagreement among correct links increases implied precision from
3.1\% to 3.5\% for Vendor B overall, from 15.1\% to 17.0\% for its
individual-level pairs, and from 13.1\% to 14.8\% for Vendor A.
The substantive interpretation is therefore unchanged across these
measurement-error scenarios.

\begin{table}[H]
\centering
\caption{Sensitivity of implied identity-level precision to demographic
measurement error among correct links, holding wrong-person demographic
agreement at $b=10\%$.}
\label{tab:measurement_error}
\begin{tabular}{lccc}
\toprule
\shortstack[l]{Disagreement among\\correct links ($e$)}
& \shortstack{Vendor B:\\all pairs}
& \shortstack{Vendor B:\\individual-level pairs}
& Vendor A \\
\midrule
0\%  & 3.1\% & 15.1\% & 13.1\% \\
1\%  & 3.2\% & 15.3\% & 13.3\% \\
5\%  & 3.3\% & 16.0\% & 13.9\% \\
10\% & 3.5\% & 17.0\% & 14.8\% \\
\bottomrule
\end{tabular}
\end{table}

\begin{figure}[H]
\centering
\includegraphics[width=\textwidth]{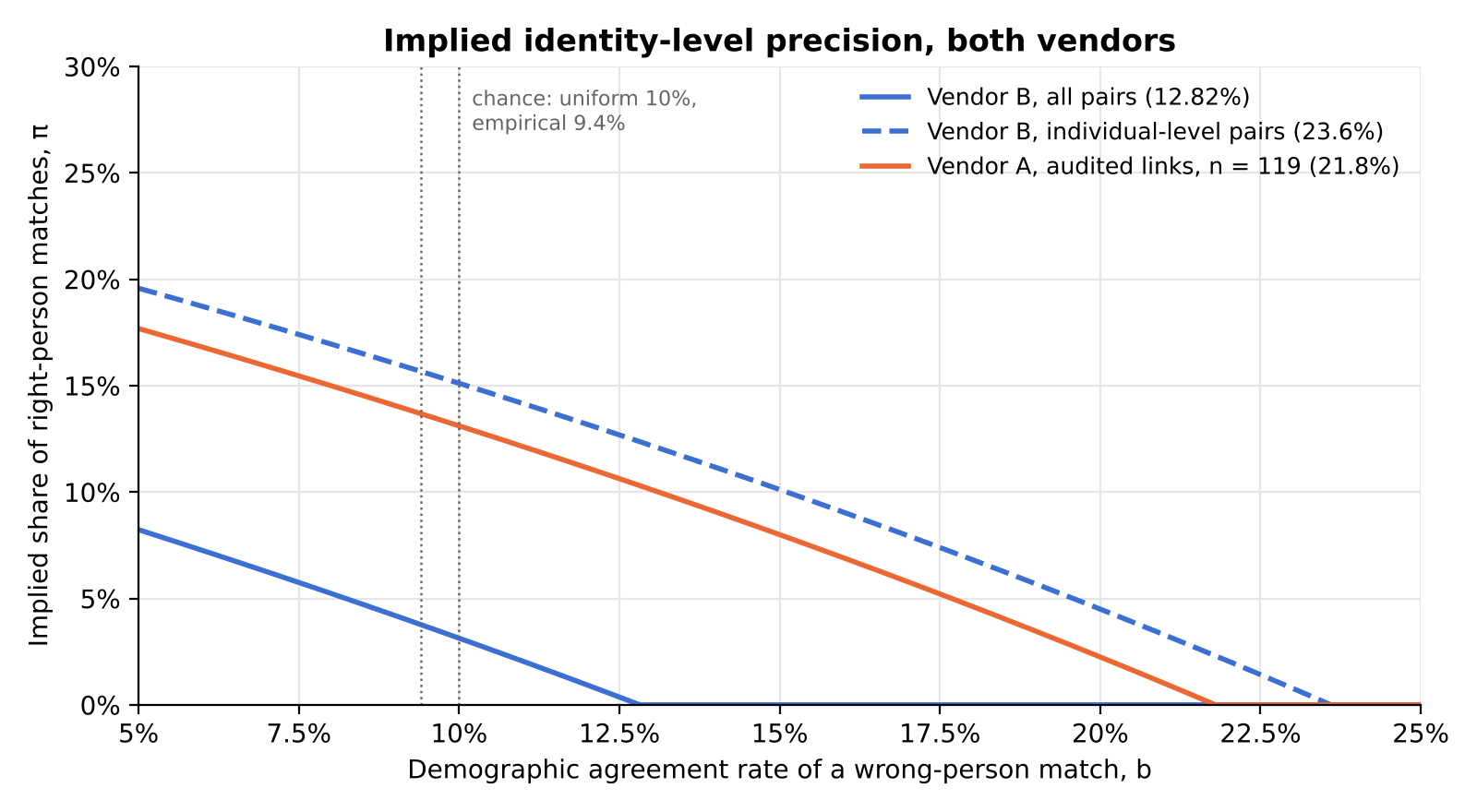}
\caption{Implied identity-level precision, both vendors. Identity-level precision $\pi$ implied by the mixture model in Equation~\eqref{eq:mixture} as a function of the demographic agreement rate $b$ of a wrong-person match, for Vendor B (all pairs and individual-level pairs) and Vendor A (audited links). Curves show point estimates. We report no intervals: a sampling interval would hold $b$ fixed. Vertical lines mark the uniform (10\%) and empirical (9.4\%) chance rates. The curves assume that correct links always agree on age bracket and gender ($e=0$); Table~\ref{tab:measurement_error} relaxes this assumption, which does not change the value of $b$ at which each curve reaches zero.}
\label{fig:overlap}
\end{figure}

Figure~\ref{fig:overlap} maps these implied precisions across a range of $b$ for both vendors (assuming $e=0$).  Wrong-person matches need not agree on demographics at the 10\% chance rate: when the wrong person is a household member, agreement may be pulled below chance (a male and a female partner share an age bracket but not a gender; parents and children usually share neither) or above chance (same-sex couples, adult siblings, multi-family homes).\footnote{Whether household composition pulls $b$ above or below 10\% is uncertain, which is why we vary $b$ rather than fix it; see Section~\ref{sec:structure} for the many-to-one structure of the analysis set.}
Implied precision falls as $b$ rises, and vice versa. An empirical baseline is available: pairing the survey panel's age and gender distribution at random with that of the voter file gives $b = 9.1$--$9.4\%$, because the panel skews young and the file old, so the uniform 10\% is slightly conservative (at $b = 9.4\%$, $\pi$ rises from 3.1\% to 3.8\% for Vendor B, and from 13.1\% to 13.7\% for Vendor A). For Vendor B's
individual-level pairs, precision reaches zero at $b=23.6\%$;
for Vendor A's audited links, at $b=21.8\%$. The pooled
Vendor B curve reaches zero at $b=12.82\%$, but includes
household-level associations and should not be interpreted
as the result for individual-level links alone. These
zero crossings identify the assumed wrong-person agreement
rates under which the model would require no correct
identities; they do not establish that those assumptions hold.

Because match rate and match quality enter effective reach multiplicatively (Section~1), a higher match rate cannot offset low quality: it enlarges the activated audience without changing its composition. At the agreement levels observed here, most of any activated audience would consist of people who were not on the submitted list, whatever the match rate.

The illustration in Section~1 shows why a higher match rate does not, by itself, establish greater correctly linked activation: the result depends on both match rate and match quality.\footnote{Our empirical audit evaluates returned record--device links, not campaign delivery. Incorrect links may connect people who are both on the submitted list, and delivery may further depend on deduplication, platform processing, and impression allocation. The observed agreement rates therefore do not directly estimate the fraction of an activated audience outside the submitted list or the fraction of impressions delivered to unintended recipients.}

\section{Conclusion}

The results in this paper should give pause to anyone who buys, sells, or evaluates identity resolution on the strength of a match rate alone. Match rate measures if submitted identifiers can be linked to other identifiers, but it does not establish if those returned identifiers are actually the same as person as the submitted identifiers. Commonly buyers have no possible method of independently evaluating the quality of those matches, thus a higher reported match rate may not actually mean more person-specific resolution but rather just lower precision among the matched identifiers.

Using data collected in 2019--2020, we assess demographic agreement in links returned by two leading identity-resolution vendors, comparing voter-file attributes with survey self-reports associated with the linked devices. The two exercises use different sampling frames and matching directions, but both reveal substantial demographic disagreement. Agreement is a diagnostic of link quality rather than direct identity verification; our identity-precision calculations are conditional on explicit measurement assumptions.  Under the stated assumptions, the implied identity-level precision is low enough to raise substantial concerns for applications that require links to identify the intended individual rather than another household member or an unrelated person.

Our measurement---one of the first of its kind, to our knowledge---finds that even among pairs Vendor B labels as individual-level matches, age bracket and gender agree in only 23.6\% of pairs. For Vendor A, the corresponding agreement is 21.8\% across 119 audited links. Under the stated mixture-model assumptions, these agreement rates imply conditional identity-level precisions of approximately 15\% and 13\%, respectively.

Findings from data that is now several years old does not confirm the performance of current identity-resolution products. But, regardless of how the match quality has changed in the last few years, our paper illustrates a timeless principle: match quality is important in addition to match rate. Incorrect identity links can compromise the campaign, especially if it includes person-specific: targeting, suppression, attribution, and other applications. 

These historical findings do not establish the performance of current products. They nevertheless illustrate a persistent measurement problem: match rate alone cannot establish match quality. Incorrect identity links can compromise person-specific targeting, suppression, attribution, and other applications.

Buyers should therefore request validation of match quality as well as match rate. Vendors should clearly define individual-level versus household-level associations, the method of validation, and the population being audited. Campaign-level aggregate numbers can appear reasonable even when many of the underlying individual-level matches are incorrect. A high match rate should therefore not be interpreted as evidence of high-quality identity resolution without corresponding evidence of match accuracy.

In closing, we note that today, this problem is just as relevant as ever, given the rise of agentic commerce and the new avenues for individual-level micro-targeting associated with these platforms.

\bibliography{references}

@article{fu2014household,
  author  = {Fu, Zhichun and Christen, Peter and Boot, Mac},
  title   = {Automatic Cleaning and Linking of Historical Census Data Using Household Information},
  journal = {International Journal of Humanities and Arts Computing},
  year    = {2014},
  volume  = {8},
  number  = {2},
  pages   = {204--225},
  doi     = {10.3366/ijhac.2014.0130},
  url     = {https://www.euppublishing.com/doi/10.3366/ijhac.2014.0130}
}

@article{ansolabehere2012validation,
  title={Validation: What big data reveal about survey misreporting and the real electorate},
  author={Ansolabehere, Stephen and Hersh, Eitan},
  journal={Political Analysis},
  volume={20},
  number={4},
  pages={437--459},
  year={2012},
  publisher={Cambridge University Press}
}

@book{hersh2015hacking,
  author    = {Hersh, Eitan D.},
  title     = {Hacking the Electorate: How Campaigns Perceive Voters},
  publisher = {Cambridge University Press},
  address   = {New York},
  year      = {2015}
}

@misc{yahoo_matchrate,
  author       = {Ray, Emily},
  title        = {Match Rate Alone Isn't Enough---Here's Why Reach Rate Matters},
  howpublished = {Yahoo Advertising Platform Insights (blog)},
  url          = {https://www.yahooinc.com/blog/match-rate-alone-isnt-enough--heres-why-reach-rate-matters},
  year         = {2025},
  note         = {Accessed September 2026}
}

@misc{liveramp_capi,
  author       = {White, Sam},
  title        = {Powering Better {Conversions API} ({CAPI}) Results with {RampID}},
  howpublished = {LiveRamp blog},
  url          = {https://liveramp.com/blog/powering-better-conversions-api-capi-results-with-rampid},
  year         = {2025},
  note         = {Accessed September 2026}
}

@misc{adexchanger_liveramp,
  author       = {Schiff, Allison},
  title        = {{LiveRamp} Launches Identity Resolution for First-Party Data},
  howpublished = {AdExchanger},
  url          = {https://www.adexchanger.com/data-exchanges/liveramp-launches-identity-resolution-for-first-party-data/},
  year         = {2021},
  note         = {Accessed September 2026}
}

\end{document}